\documentclass[sigconf]{acmart}

\usepackage{multirow}
\usepackage{graphicx}
\usepackage[caption=false]{subfig}
\usepackage[export]{adjustbox} 
\usepackage{xcolor}
\usepackage{fancybox}
\usepackage{framed}
\usepackage{minibox}
\usepackage{svg}
\usepackage{stfloats}

\usepackage{soul}
\newcommand{\ctext}[3][RGB]{%
  \begingroup
  \definecolor{hlcolor}{#1}{#2}\sethlcolor{hlcolor}%
  \hl{#3}%
  \endgroup
}

\AtBeginDocument{%
  \providecommand\BibTeX{{%
    \normalfont B\kern-0.5em{\scshape i\kern-0.25em b}\kern-0.8em\TeX}}}

\copyrightyear{2024}
\acmYear{2024}
\setcopyright{acmlicensed}\acmConference[ICPC '24]{32nd IEEE/ACM International Conference on Program Comprehension}{April 15--16, 2024}{Lisbon, Portugal}
\acmBooktitle{32nd IEEE/ACM International Conference on Program Comprehension (ICPC '24), April 15--16, 2024, Lisbon, Portugal}
\acmDOI{10.1145/3643916.3644413}
\acmISBN{979-8-4007-0586-1/24/04}

\begin{document}



\title{Rationale Dataset and Analysis 
 for the Commit Messages of the Linux Kernel Out-of-Memory Killer}


\author{Mouna Dhaouadi}
\affiliation{%
  \institution{DIRO, Universit\'{e} de Montr\'{e}al}
  \city{Montr\'{e}al}
  \country{Canada}}
\email{mouna.dhaouadi@umontreal.ca}

\author{Bentley James Oakes}
\affiliation{%
  \institution{GIGL, Polytechnique Montr\'{e}al}
    \city{Montr\'{e}al}
  \country{Canada}}
\email{bentley.oakes@polymtl.ca}

\author{Michalis Famelis}
\affiliation{%
  \institution{DIRO, Universit\'{e} de Montr\'{e}al}
  \city{Montr\'{e}al}
  \country{Canada}}
\email{michalis.famelis@umontreal.ca}

\renewcommand{\shortauthors}{Dhaouadi, et al.}

\begin{abstract}

Code commit messages can contain useful information on \textit{why} a developer has made a change. However, the presence and structure of rationale in real-world code commit messages is not well studied. Here, we detail the creation of a labelled dataset to analyze the code commit messages of the Linux Kernel Out-Of-Memory Killer component. We study aspects of rationale information, such as presence, temporal evolution, and structure. We find that 98.9\% of commits in our dataset contain sentences with rationale information, and that experienced developers report rationale in about 60\% of the sentences in their commits. 
We report on the challenges we faced and provide examples for our labelling.


\end{abstract}




\maketitle

\section{Introduction}
\label{sec:intro}

To be effective in evolving software systems, developers who desire to propose code changes have to deeply understand the system and its behavior. Developing this understanding is not easy, yet it is crucial to grasp the \textit{rationale} behind the decisions that shaped the system to its current state.
Communities of software developers may put in place norms to encourage documenting this rationale, to describe \textit{why} each change was made. In modern software development, such information is often contained in the \textit{commit message} that accompanies a proposed code change that is submitted in a shared version control repository, such as Git. 

Although several researchers have discussed the importance of rationale~\cite{burge2008rationale}, and proposed various approaches for structuring and extracting it~\cite{kleebaumContinuousRationaleIdentification, dhaouadi2022end, bhatAutomaticExtractionDesign2017, rogers2012exploring, liangLearningWhysDiscovering2012, vandervenMakingRightDecision2013, dhaouadi2023towards}, 
%
%
there are scant details in the literature about the characteristics of rationale in real-world systems. In fact, only few researchers have attempted to develop a deep understanding of developers' rationale in Open Source Software (OSS), e.g., by studying chat messages or email archives~\cite{alkadhiHowDevelopersDiscuss2018, sharmaExtractingRationaleOpen2021}. 
%
%
%
%
%
%
To the best of our knowledge, there is \textit{no prior work studying developer's rationale in the code commit messages of open source projects, specifically their structure}. We expect that the results of such studies could be used a) to better understand the characteristics of rationale (such as its presence, structure and evolution over a project's lifetime), 
b) as first steps towards setting rationale documentation guidelines to improve commit messages to justify the commits for OSS commits in the future, which would reduce the developer's need to access an issue tracker (avoiding inconsistencies or traceability errors), and c) as baselines for automated solutions (e.g. GitHub pull request bots) to identify rationale in software. 



To fill this gap,  we propose to study developers' rationale in the commit messages of an open source project. Our main research questions is: \textit{What are the characteristics (presence, impact factors, evolution, structure) of how rationale information appears in collaborative open-source commit messages?} Specifically, we report on our creation of an \textit{annotated, high-quality rationale dataset} for the Out-of-Memory Killer (OOM-Killer) component of the Linux kernel project. We have previously argued that this project is well suited for studying rationale~\cite{dhaouadi2022end}. In this work, we systematically annotate the content of these commit messages, using a categorization of sentences as \textit{Decision}, \textit{Rationale} and \textit{Supporting Facts}~\cite{dhaouadi2023towards}. We then analyze quantitatively the resulting dataset to characterize rationale in the OOM-Killer subsystem. Specifically, we follow the empirical standards of a repository mining study~\cite{ralph2020empirical}.

We previously proposed \textit{Kantara}~\cite{dhaouadi2022end}, an end-to-end automatic rationale extraction and management pipeline, and evaluated it on some example commits from the Linux OOM-Killer component. Our then-partial labelling of the OOM-Killer dataset is completed here, as a stepping stone to future work on automatic extraction~\cite{dhaouadi2022extraction}. We also described how this rationale information can be represented using an ontologically-based knowledge graph, along with some initial reporting and visualization functionalities of \textit{Kantara}~\cite{dhaouadi2023towards}. 


The contributions of this paper are twofold: 
1) a high-quality dataset of labelled commits of the OOM-Killer component, and 2) an analysis of the dataset of seven research questions (see Section~\ref{sec:dataset_analysis}) that touch on topics ranging from the frequency of the presence of rationale, the factors that impact it, its temporal evolution, to the structure of commit messages. 
The dataset and the source code used to perform our analysis are publicly available\footnote{\scriptsize
\url{https://zenodo.org/records/10063089}
}.

The remainder of this paper is organized as follows. We present the OOM-Killer subsystem in Section~\ref{sec:linux}, and discuss our work on creating the labelled dataset of the OOM commits in Section~\ref{sec:dataset_creation}. Then, we describe the obtained dataset and analyze it in Section~\ref{sec:dataset_des_analysis}. 
We introduce the threats to validity we encountered in Section~\ref{sec:threats},
overview  related work in Section~\ref{sec:related}, and conclude in Section~\ref{sec:conclusion}.

\section{Linux Out-Of-Memory Killer Subsystem}
\label{sec:linux}

The Linux kernel is a large open-source project that is being continuously developed collaboratively since 1991. Its main collaboration channel is the Linux Kernel Mailing List (LKML), which contains code patches and discussions. Since 2005, code patches for the Linux kernel have been structured as \textit{Git commits}. 
Table~\ref{tab:coloured_commit_no_overlap} shows an example of a commit message. Developers are encouraged to explain the motivation for the commit and explain its impact on the kernel in the commit message\footnote{\scriptsize \url{https://www.kernel.org/doc/html/latest/process/submitting-patches.html}.}.
This community practice makes the Linux kernel commits a rich repository of rationale information~\cite{dhaouadi2022end}. 



The \textit{Out-Of-Memory Killer (OOM-Killer) subsystem} is a Linux module~\cite{bovet2005understanding} that is in charge of freeing up the memory to prevent a system crash when all the available memory has already been allocated. It embodies one approach to handling OOM problems~\cite{huang2016evolutionary} by using a set of heuristics to select a task (the \textit{OOM victim}) and ``killing'' it, i.e., forcing it to terminate. The victim is thus forced to release its memory and exit. We have previously argued that this OOM-Killer component is a particularly good source of rationale, as its commit messages reveal interesting decisions about the best  selection strategy of the OOM victim~\cite{dhaouadi2022end}. 

Thus, we have applied purposeful sampling~\cite{suri2011purposeful}. We have selected a component (the OOM-Killer) with high-quality justification in commit messages as a critical case, i.e., a case where rationale of high quality is expected to be present and identifiable. Absence of rationale in this case would signify deeper problems for rationale identification in general.


\section{Dataset Creation}
\label{sec:dataset_creation}

This section discusses labelling the rationale information of all commits of the OOM subsystem, as well as the final data set structure.

\subsection{Commit Pre-processing}
We obtained the initial set of commits downloading the available commit history for the oom$\_$kill.c file which contains the C source code of the OOM Killer module\footnote{\scriptsize \url{https://github.com/torvalds/linux/commits/master/mm/oom_kill.c}, accessed on 12/01/2023.}. This initial set contained 418 commits since the Linux development moved to Git on 2005-04-16 up until 2022-09-27 (the commit date of the latest commit we pulled). We excluded Git merge commits, as they do not typically contain informative commit messages.
In fact, in this particular module, we have 13
merge commits that restate previous commits (e.g., ''Pull updates from X"\footnote{\scriptsize \url{https://github.com/torvalds/linux/commit/35ce8ae9ae2e471f92759f9d6880eab42cc1c3b6}}), contain no description of the changes (e.g., enumerate a list of patches\footnote{\scriptsize\url{https://github.com/torvalds/linux/commit/512b7931ad0561ffe14265f9ff554a3c081b476b}}) or summarize the changes in a non-informative manner (e.g., ``updates''\footnote{\scriptsize\url{https://github.com/torvalds/linux/commit/28e92f990337b8b4c5fdec47667f8b96089c503e}}). Therefore, removing them does not present a threat to validity. Only one merge commit contains informative description as it fixes some conflicts\footnote{\scriptsize \url{https://github.com/torvalds/linux/commit/2b828925652340277a889cbc11b2d0637f7cdaf7}}. Although removing this specific commit presents a small threat, we chose to do so for consistency reasons.

We then performed pre-processing of the messages of the remaining 404 non-merge commits. For each message, we removed the meta-data at the end of the message, such as the tags \textit{Signed-off-by}  and \textit{Suggested-by}, that are not relevant for the study of rationale. We also removed URLs, references to other resources, and call traces using regular expressions. We then split the message to sentences, keeping only sentences with: a) more than three characters, and b) that are not source code. We used heuristics to detect whether a sentence is source code, like the existence of keywords or symbols such as \textit{git}, \textit{\$cd} or \textit{\$echo}. We chose these keywords by manually investigating the data. 

Since the data was not properly formatted, the pre-processing step was challenging and only partially successful. For instance, entries such as \textsl{``BUG: scheduling while atomic: rsyslogd/1422/0x00000002 INFO: lockdep is turned off''}\footnote{\scriptsize \url{https://api.github.com/repos/torvalds/linux/git/commits/b52723c5607f7684c2c0c075f86f86da0d7fb6d0}} were not removed. To overcome this limitation, we will continue cleaning the data during the labelling, using the \textit{Inapplicable} label (see Section~\ref{sec:codebook}). The pre-processed sentences could be also be corrected manually but that would be time consuming and error-prone.

\subsection{Sentence Labelling Procedure}
\subsubsection{Piloting.} 
 We continue the categorization of \textit{Decision}, \textit{Rationale} and \textit{Supporting Facts} from~\cite{dhaouadi2023towards}, as shown in Table~\ref{tab:codebook}. We note that a sentence can have multiple labels, or no labels. 
 To develop a shared understanding regarding the meaning of the labels we will be using, we performed five iterations of piloting rounds and consolidation meetings during which the three annotators\footnote{\scriptsize A PhD student, a post-doctoral researcher, and a professor.} independently annotated 33 randomly-chosen commits in total.  We elaborated a common protocol to consistently label the dataset based on the diverse cases we encountered during our piloting rounds.

\begin{table}[t]
\small
  \caption{Labelling codebook}
  \label{tab:codebook}
  \begin{tabular}{p{2.2cm}p{5.5cm}}
    \toprule
    Label & Meaning\\
    \midrule
    Decision& An action or a change that has been made, including a description of the patch behaviour \\
     Rationale& Reason for a decision or value judgment
     \\
   Supporting Facts & A narration of facts used to support a  decision \\
     Inapplicable &   Pre-processing error or bad sentences\\
     & (i.e., does not contain English sentences) \\
  \bottomrule
\end{tabular}
\vspace{-0.15in}
\end{table}

\subsubsection{Codebook.}
\label{sec:codebook}
As shown in Table~\ref{tab:codebook}, a \textit{Decision} provides information about the state of the system
after the patch is applied, i.e., it refers to the system’s
future state. 
\textit{Rationale} is the reason why a decision is taken,
such as a value judgement about undesirable behavior.
\textit{Supporting Facts} are bits of information in a sentence
where a developer discusses the currently existing state of
the system, at the moment before they propose a change.  
The \textit{Inapplicable} label is used by the annotators to identify noise in the data that was not successfully filtered by the pre-processing step due to the lack of uniform formatting and the presence of (pseudo-)code.

Table~\ref{tab:coloured_commit_no_overlap} reproduces a commit from the dataset, along with a classification for each sentence. 
The first sentence (the summary phrase of the commit) is labelled as a \textit{Decision} as it states the patch’s change. 
The second sentence is labelled as a \textit{Supporting Facts} as it presents the currently existing state of the system, and the third sentence is labelled as \textit{Rationale} as it motivates this commit. 



\begin{table*}[t]
\small
    \centering
     \caption{An example commit with labelled sentences from our dataset}
    \begin{tabular}{p{14cm}|c}
       \textbf{Sentence}  & \textbf{Labelling} \\\hline
      \ctext[RGB]{173, 216, 230}{mm, oom: introduce independent oom killer ratelimit state} & Decision\\

 \ctext[RGB]{255, 250, 205}{printk\_ratelimit() uses the global ratelimit state for all printks} & Supporting Facts\\
 
\ctext[RGB]{244, 194, 194}{The
oom killer should not be subjected to this state just because another
subsystem or driver may be flooding the kernel log} & Rationale\\

\ctext[RGB]{173, 216, 230}{This patch introduces printk ratelimiting specifically for the oom killer.} & Decision \\

    \end{tabular}
    \label{tab:coloured_commit_no_overlap}
    
\end{table*}


The codebook we adopted is data-driven; we came to it after several discussions and consolidation meetings.
However, this categorization is preliminary and used as a first-order classification. For example, although we labelled the sentence \textsl{``A future optimization could be to put sched\_entity and sched\_rt\_entity into a union.''}\footnote{\scriptsize \url{https://api.github.com/repos/torvalds/linux/git/commits/fa717060f1ab7eb6570f2fb49136f838fc9195a9}} as \textit{Decision}, it could also fit under a \textit{Suggestion} category. We plan to extend this codebook in the future to include other components of commit message rationale (e.g. Goal, Need, Benefit, etc.~\cite{alsafwanDevelopersNeedRationale2022}).

\subsubsection{Protocol.} 
\label{sec:protocol}
To systematically create the dataset, we developed this protocol for annotating the commit message sentences:

\begin{enumerate}

\item We do not separate a sentence from its context; when in doubt, we look at the patch code to better understand it.

\item Past changes count as \textit{Supporting Facts}. \textit{Decision} and \textit{Rationale}, however, usually concern the present (current commit). \textit{e.g.}, the sentence \textsl{``The reason this check was added was the thought that, since only the OOM disabling code would wait on this queue, wakeup operations could be saved when that specific consumer is known to be absent.''}\footnote{\scriptsize \url{https://api.github.com/repos/torvalds/linux/git/commits/c38f1025f2910d6183e9923d4b4d5804474b50c5}} counts as \textit{Supporting Facts}. In contrary, the sentence \textsl{``It's better to extract this to a helper function to remove all the confusion as to its semantics.''}\footnote{\scriptsize \url{https://api.github.com/repos/torvalds/linux/git/commits/309ed882508cc471320ff79265e7340774d6746c	}} counts as \textit{Decision} and \textit{Rationale}.

\item Future intent counts as rationale.  \textit{e.g.}, the sentence \textsl{``Moreover, this will make later patch in the series easier to review.''}\footnote{\scriptsize \url{https://api.github.com/repos/torvalds/linux/git/commits/7ebffa45551fe7db86a2b32bf586f124ef484e6e}} counts as \textit{Rationale}.

\item We consider terse value judgment language (\textit{e.g.}, ``fix'' or ``cleanup'') to imply the presence of rationale and descriptions of decisions, even if it is low quality. \textit{e.g.}, the sentences \textsl{``mm/oom\_kill.c: fix vm\_oom\_kill\_table[] ifdeffery.''}\footnote{\scriptsize \url{https://api.github.com/repos/torvalds/linux/git/commits/a19cad0691597eb79c123b8a19a9faba5ab7d90e}} and \textsl{``Unify it.''}\footnote{\scriptsize \url{https://api.github.com/repos/torvalds/linux/git/commits/ab290adbaf8f46770f014ea87968de5baca29c30}} count both as \textit{Decision} and \textit{Rationale}.

\item It is possible that no label is applicable.
\textit{e.g.}, the sentence \textsl{``mm/ mmu\_notifier: contextual information for event triggering invalidation.''}\footnote{\scriptsize \url{https://api.github.com/repos/torvalds/linux/git/commits/6f4f13e8d9e27cefd2cd88dd4fd80aa6d68b9131}} does not fit any of the three categories. 

\item Examples can also be labelled with one of the categories. \textit{e.g.}, the sentence \textsl{``In the following example, abuse\_the\_ram is the name of a program that attempts to iteratively allocate all available memory until it is stopped by force.''}\footnote{\scriptsize \url{https://api.github.com/repos/torvalds/linux/git/commits/8ac3f8fe91a2119522a73fbc41d354057054e6ed	}} counts as \textit{Supporting Facts}. However, the sentence \textsl{``The value is added directly into the badness() score so a value of -500, for example, means to discount 50\% of its memory consumption in comparison to other tasks either on the system, bound to the mempolicy, in the cpuset, or sharing the same memory controller.''}\footnote{\scriptsize \url{https://api.github.com/repos/torvalds/linux/git/commits/a63d83f427fbce97a6cea0db2e64b0eb8435cd10}} counts as \textit{Decision}.

\item If a sentence was mistakenly cut during pre-processing,  we label all the parts with the same labels. \textit{e.g.}, these two parts of the same sentence: \textsl{``oom\_reaper used to rely on the oom\_lock since e2fe14564d33 ("oom\_reaper''} and \textsl{``close race with exiting task").''}\footnote{\scriptsize \url{https://api.github.com/repos/torvalds/linux/git/commits/af5679fbc669f31f7ebd0d473bca76c24c07de30	}} should be labelled the same even though pre-processing produced them as separate items.

\item If pre-processing has produced an item that mixes code (\textit{e.g.}, log or trace or source code) with a valid English sentence (\textit{e.g.}, not heading, for example: "current message"), we ignore the code and label the item based on the sentence. e.g., when labelling the item \textsl{``oom-kill: constraint=CONSTRAINT\_NONE, nodemask=(null), cpuset=/,
  mems\_allowed=0-1, task=panic, pid=10737, uid=0 An admin can easily get the full oom context at a single line which makes parsing much easier.''}\footnote{\scriptsize \url{https://api.github.com/repos/torvalds/linux/git/commits/ef8444ea01d7442652f8e1b8a8b94278cb57eafd	}}, we disregard the non-English part and apply a label based only on the part starting with \textsl{``An admin can easily...''}.

\end{enumerate}

\subsubsection{Labelling.}

We conducted the labelling of the 366 remaining commits (i.e, excluding those used for the piloting) in batches, over a period of six months. In all, we annotated 2333 sentences, where sentences could have multiple labels. While labelling independently, we held regular meetings to discuss problematic sentences and resolve discrepancies. The purpose of the meetings was to agree on how to remove any potential misunderstanding or confusion regarding the commit messages. For sentences where we could not establish complete agreement or that contained ambiguous language, we assigned labels by taking the union of all opinions.

 \begin{table}[t]
 \footnotesize
     \centering
     \caption{Inter-rater reliability table}
     \begin{tabular}{|c|c|c|c|c|}
     \hline
      \multirow{2}{*}{Batch}&\multirow{2}{*}{Size}  & \multicolumn{3}{c|}{Fleiss Kappa}  \\ 
       \cline{3-5} 
          &   & Decision & Rationale & Supporting Facts  \\
        \hline
         1 &  20  &  0.596  & 0.514   &    0.489  \\
         2 &  20  &  0.733  & 0.487   &   0.596  \\
         3 &  20  & 1.0  &  0.933   &   1.0  \\
         4 &  40  & 0.899  & 0.929   &   1.0  \\
         5 &  20  &  0.928 & 0.796   &  0.738   \\
         6 &  20  & 0.861  & 0.80   &   0.40  \\
         7 &  20  & 1.0  & 0.853   &    0.863 \\
         8 &  20  & 0.865  & 0.498   &   0.666  \\
         9 &  20  & 0.644  & 0.261   &   0.799  \\
        10 &  20  & 0.286  & 0.20   &   0.280  \\
        11 &  20  & 0.796  & 0.865   &  0.598   \\
        12 &  20  &  0.777  & 0.707   &   0.665  \\
        13 &  20  & 0.925  & 0.498   &    0.850  \\
        14 &  20  & 0.569  & 0.343   &   0.603  \\
        15 &  20  & 1.0  & 0.932  &     0.731 \\
        16 &  20  &  0.918  & 0.555   &  0.859   \\
        17 &  20  &  0.822  &  0.775   &   0.733  \\
        18 &  6   &  0.593  & 0.326   &   0.175  \\
         \hline
        \textbf{All} & \textbf{366} & \textbf{0.748} &  \textbf{ 0.603}& \textbf{0.648}  \\ 
        \hline
     \end{tabular}
     \label{tab:batches}
     \vspace{-0.15in}
 \end{table}


\begin{table*}[tbh]
\footnotesize
    \centering
     \caption{Example of the dataset structure with one entry}
    \label{tab:one_entry}
   \begin{tabular}{p{2.55cm}p{14cm}}
    \toprule
    Column & Value\\
    \midrule
    commit number  &  4 \\
    commit ID   &  C\_kwDOACN7MtoAKGExOWNhZDA2OTE1OTdlYjc5YzEyM2I4YTE5YTlmYWJhNWFiN2Q5MGU	 \\
 author name   &  Andrew Morton  \\
 committer name  &  akpm  \\
 message	   & mm/oom\_kill.c: fix vm\_oom\_kill\_table[] ifdeffery arm allnoconfig: mm/oom\_kill.c:60:25: warning: 'vm\_oom\_kill\_table' defined but not used [-Wunused-variable] 60 | static struct ctl\_table vm\_oom\_kill\_table[] =  Cc: Luis Chamberlain <mcgrof@kernel.org> Signed-off-by: Andrew Morton <akpm@linux-foundation.org> \\
 URL  &   {\url{https://api.github.com/repos/torvalds/linux/git/commits/a19cad0691597eb79c123b8a19a9faba5ab7d90e}} \\
 message\_preprocessed  &   mm/oom\_kill.c: fix vm\_oom\_kill\_table[] ifdeffery \\
 Decision  &  yes\\
 Rationale  &  yes\\ 
 Supporting Facts   & no \\
  \bottomrule
\end{tabular}

\end{table*}

To compute inter-rater agreement, we use Fleiss Kappa 
\cite{fleiss1971measuring} 
since we are in the presence of more than two annotators. Table~\ref{tab:batches} shows the number of batches, their sizes (number of commits) and the Fleiss Kappa per category. Fleiss Kappa was computed after the consolidation meetings. As shown in Table~\ref{tab:batches}, there is generally a stronger consensus about the \textit{Decision} and \textit{Supporting Facts} categories than the \textit{Rationale} category. Despite the low kappa values for individual categories in some batches (e.g, batch 6, 9 or 10),  overall the kappas are  considered \textit{good} ($>0.6$) for the three categories. This indicates strong agreement considering the subjective nature of rationale~\cite{burge2008rationale}, and demonstrates the high quality of the dataset.

The main challenge we faced during the labelling was understanding the commit messages, such as due to  implicit or ambiguous language.  As some commit messages are written by (presumed) non-native authors, it was hard to read the tense of the sentences, especially past/future tenses.
For example, it is hard to distinguish whether \textsl{``The oom\_reaper end then simply retries if there is at least one notifier which couldn't make any progress in !blockable mode''\footnote{\scriptsize \url{https://api.github.com/repos/torvalds/linux/git/commits/93065ac753e4443840a057bfef4be71ec766fde9}}} is a statement before or after the patch.
Most commits also needed  technical understanding. 
Although we tried our best to interpret what the sentence contains, it is still possible that we
misunderstood some of the commit messages. 
To mitigate any potential bias, we took an inclusive labelling approach.
We consider as future work involving Linux developers to obtain a more reliable labelling and a finer-grained analysis. 

\subsection{Dataset Structure}

After  removing 99 sentences that were flagged as \textit{Inapplicable} by at least two annotators, we end up with 2234 sentences. We consider the final classification of these sentences as the union of the labels of the three annotators. We use a comma separated values (.CSV) tabular format where each
entry (line) refers to a sentence for easy analysis.  An example of the dataset structure is shown in Table~\ref{tab:one_entry}.

\section{Dataset Description and Analysis}
\label{sec:dataset_des_analysis}


\subsection{Dataset Description}
\label{sec:dataset_description}

Here, we describe the distribution of sentences over the agreed-upon labels. We also overview the most used words in each category.

\subsubsection{Sentences distribution.}

\begin{figure}[t]
    \centering
    \includegraphics[width=0.2\textwidth,trim=1 2 1 1,clip]{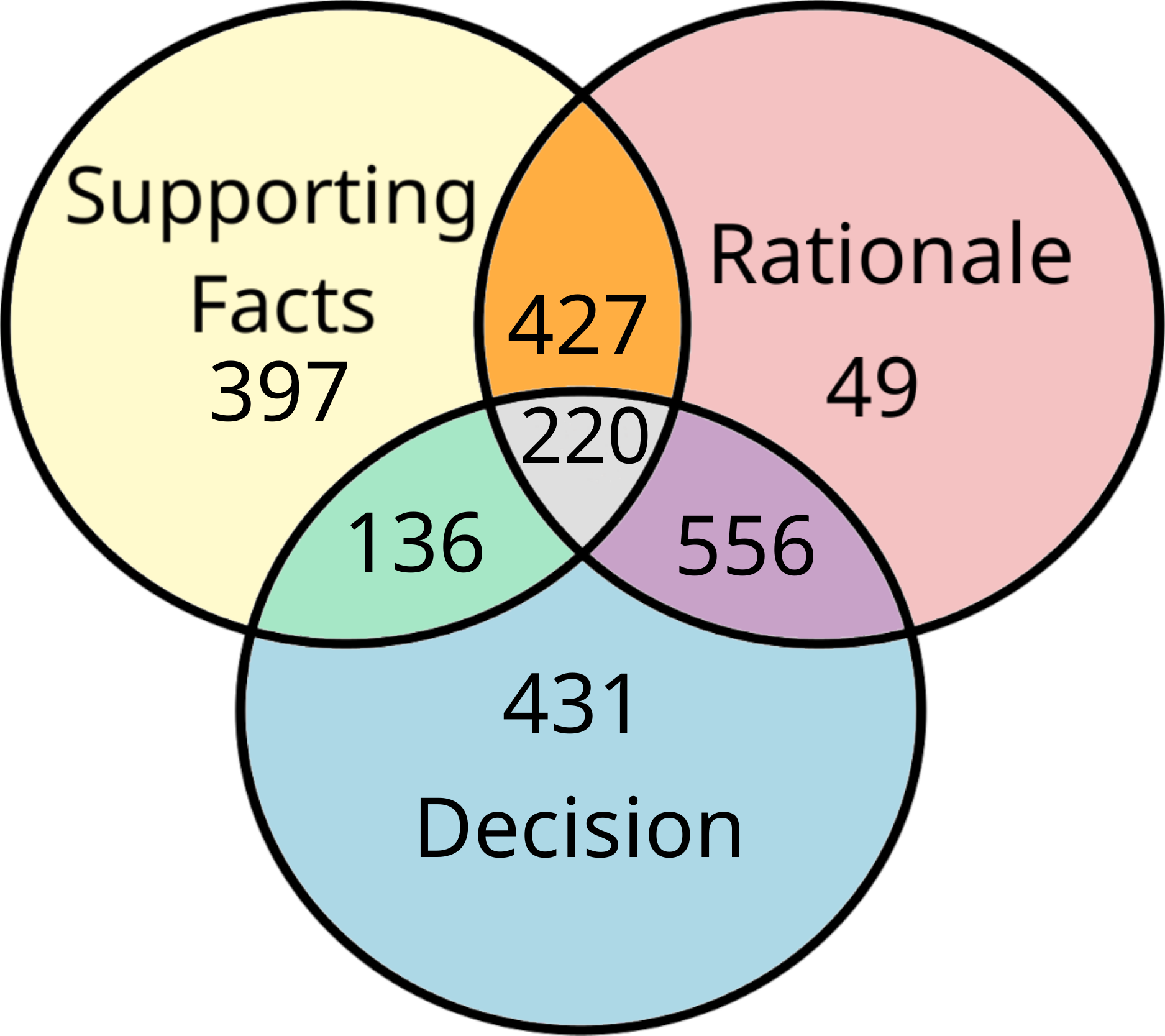}
    \caption{Distribution of the sentences in our OOM dataset}
    \label{fig:dataset_categories}
 \vspace{-0.15in}
\end{figure}

Among the 2234 sentences in the dataset, 18 did not fit in any category.  
Figure~\ref{fig:dataset_categories} shows the distribution of the remaining 2216 labelled sentences
over the identified categories
The distribution in Figure~\ref{fig:dataset_categories} shows that the three categories are not clear-cut and that there is a substantial overlap between them.  In fact, among the 366 commits in question, only two commits had all their sentences classified with only one label (one of them is the commit presented in Table~\ref{tab:coloured_commit_no_overlap}). 
This could be explained by our inclusive labelling for disagreement. We also expected sentence-level labelling to lead to multi-category classification. In the future, we will investigate further categories, and whether phrase-level labelling can reduce overlap.


Furthermore, we observe a large intersection between 
\textit{Rationale} and the other categories. One interpretation of this is that it indicates the subjective nature of rationale. Another is that \textit{rationale} is often present in the same sentence to motivate \textit{decisions} and to employ \textit{supporting facts}. We show examples of such multi-labelled sentences in Table~\ref{tab:coloured_commit_with_overlap}. In it, we show a commit from our dataset, along with
a colour-coded multi-label classification for each sentence, using the same colour scheme as in Figure~\ref{fig:dataset_categories}. As
an example of the labelling, the second sentence contains a \textit{supporting fact} (``...~now that there is a separate function for 'fullmm' flushing'') that reinforce the \textit{rationale} behind this patch (``The 'start' and 'end' arguments to tlb\_gather\_mmu() are no longer needed''). The third sentence is labelled as both \textit{Decision} and \textit{Rationale} as it  states  not only the patch’s change (``Remove the .. and update all callers'') but also the motivation behind this change, expressed as a value judgment (``unused arguments'').

\begin{table*}[]
\small
    \centering
     \caption{An example commit  with multi-labelled sentences from our dataset, using the same colour scheme as in Figure~\ref{fig:dataset_categories}.}
    \begin{tabular}{p{13cm}|c}
       \textbf{Sentence}  & \textbf{Labelling} \\\hline
       \ctext[RGB]{173, 216, 230}{tlb: mmu\_gather: Remove start/end arguments from tlb\_gather\_mmu()} & Decision\\

\ctext[RGB]{255, 174, 66}{The 'start' and 'end' arguments to tlb\_gather\_mmu() are no longer
needed now that there is a separate function for 'fullmm' flushing} &  Rationale, Supporting Facts\\

\ctext[RGB]{200, 162, 200}{Remove the unused arguments and update all callers.} & Decision, Rationale\\
 \end{tabular}
 \label{tab:coloured_commit_with_overlap}
\end{table*}

\subsubsection{Frequent words.}

To better understand how developers express themselves in a large-scale open source project, we extract the most prominent words in each of the categories. We augmented the built-in list of stop words of the wordcloud python library\footnote{\url{https://amueller.github.io/word_cloud/}} to be removed
with the following list of words:
\textit{OOM}, \textit{mm}, \textit{memory}, \textit{killer}, \textit{kernel}, \textit{victim}, \textit{task}, \textit{linux}, \textit{thread}, \textit{process}, \textit{system}, \textit{patch}, \textit{oom\_kill}, \textit{memcg}, as they have no  added value for our interpretation. We do not consider multi-labelled sentences as we are most interested in discovering the distinct characteristics of each category. 

\begin{figure}[ht]
\centering 
\captionsetup{justification=centering}
\subfloat[Decision word cloud. Most frequent words: 
 `add' 28,
 `use' 28,
 `remove' 28,
 `kill' 27,
 `tasks' 22,
 `set' 20,
 `cpuset' 20,
 `instead' 19,
 `introduce' 18,
 `check' 16 \label{fig:wordclouds-a}]{%
   \includegraphics[width=0.3\textwidth, frame]{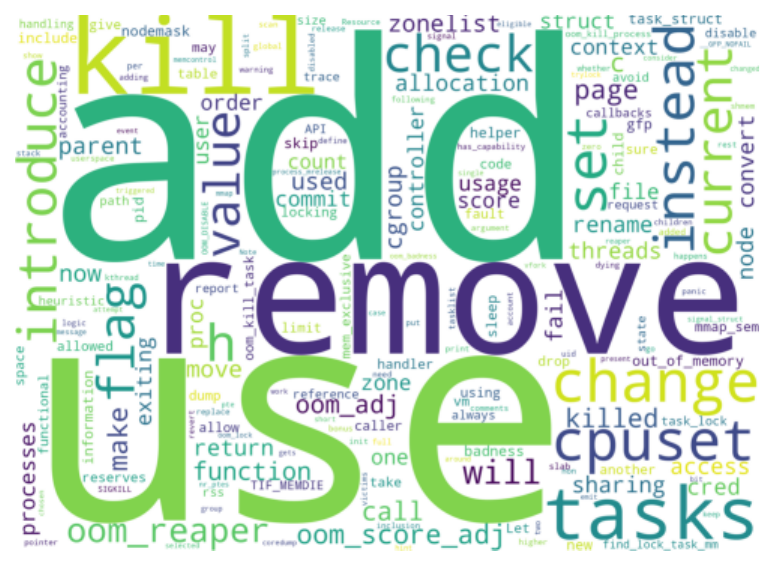}%
}

\subfloat[Rationale word cloud. Most frequent words: 
 `might': 5,
 `make' 5,
 `will' 4,
 `fixes' 4,
 `help' 4,
 `debugging' 4,
 `later' 4,
 `use' 3,
 `reduce' 3,
 `useful' 3\label{fig:wordclouds-b}]{%
  \includegraphics[width=0.3\textwidth, frame]{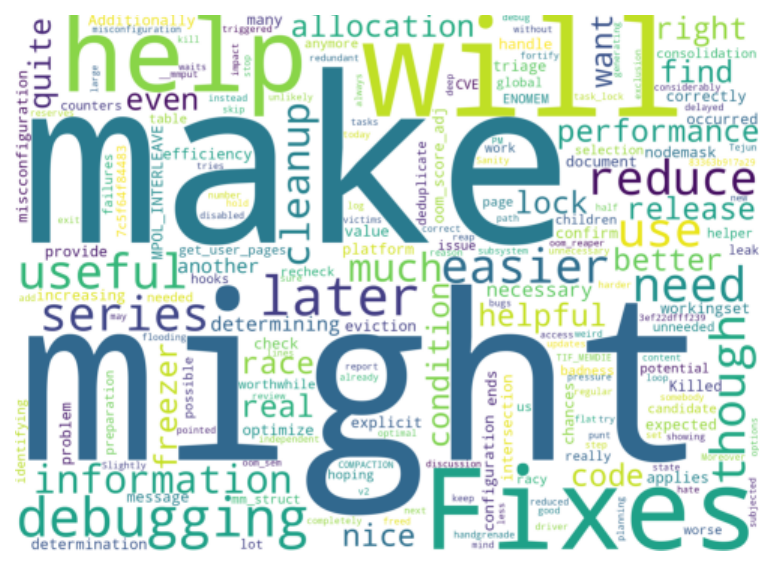}%
   
}\hfil

\subfloat[Supporting Facts word cloud. Most frequent words: `kill' 46,
 `will' 29,
 `tasks' 28,
 `node' 27,
 `killed' 26,
 `current' 26,
 `allocation' 24,
 `set' 23,
 `check' 21,
 `reaper' 20\label{fig:wordclouds-c}]{%
 \includegraphics[width=0.3\textwidth, frame]{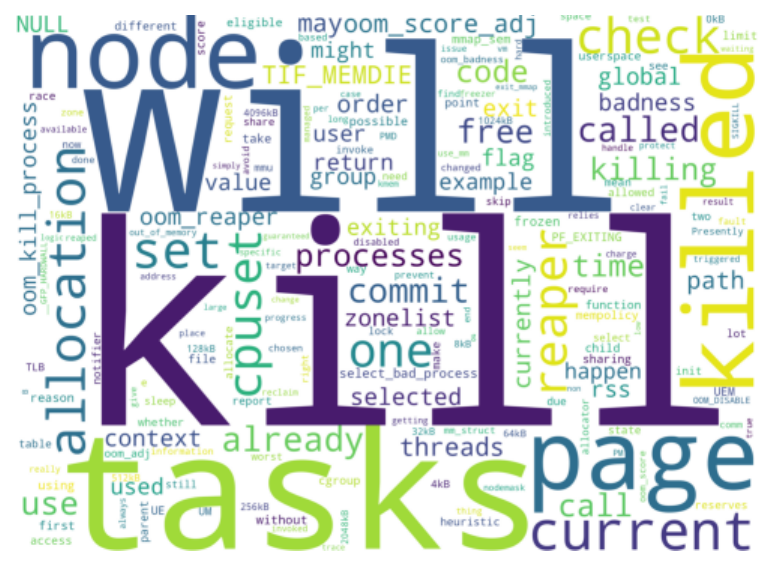}%
}\hfil

\caption{Most frequent words per category, without overlap}
\label{fig:wordclouds}
\end{figure}

In Figure~\ref{fig:wordclouds-a}, we visualize the most frequent words in the 431 sentences labelled as \textit{Decision} only. Action verbs are very common in how developers phrase their decisions (e.g, ``add'', ``remove'', ``use'', ``introduce'', ``change''). 
In Figure~\ref{fig:wordclouds-b}, we visualize the most used words in the 49 sentences labelled as \textit{Rationale} only. We notice that the (``make'', ``might'' ) are the most important. Also,   verbs that imply value judgment (e.g, ``fixes'', ``cleanup'') and positive connotations adjectives (e.g, ``useful'', ``easier'') are frequent. Finally, we can notice the reference to the future (e.g, ``will'', ``later''). This indicates that developers tend to express the future positive impact of their patches.
In Figure~\ref{fig:wordclouds-c}, we visualize the most used words in the 397 sentences labelled as \textit{Supporting Facts} only. We note the presence of context-specific words (e.g, ``kill'', ``killed'', ``tasks'', ``node''). One interpretation of this is that supporting facts are descriptions of the existing state of the system. The verb ``will'' is also frequent in this category. This indicates that when developers describe the current state of the system, they tend to mention the inevitable events, probably to motivate their changes that would avoid them.
These observations could be used in the future as a basis to propose heuristics for automatic rationale extraction in the Linux Kernel.

\subsection{Dataset Analysis}
\label{sec:dataset_analysis}

We present seven research questions (RQs) on four themes: 
(a) the presence of rationale (RQ1, RQ2), 
(b) the factors that impact it (RQ3, RQ4), 
(c) its temporal evolution (RQ5, RQ6), 
and (d) the structure of commit messages from the point of view of rationale (RQ7).


\subsubsection{Presence of Rationale.}
Here, we discuss rationale information \textit{abundance} (\textbf{RQ1}) and \textit{amount} (\textbf{RQ2}) in the commit messages.

\textbf{RQ1.  How many commits contain rationale?} To answer this question, we compute the \textit{rationale \%} as follows:\\ 

 \begin{math}
 \textit{rationale density} \% = \frac{ \text{number of commits that contain rationale} }{\text{total number of commits} }\\
 \end{math}
 
We consider that a commit contains rationale if at least one of its sentences is labelled as \textit{Rationale}. In our dataset, 98.9\% of the commits contain at least one sentence with rationale information. This suggests that rationale is almost always described.

\textbf{RQ2. How much of the commit contains rationale?}
To answer RQ2, we define the commit-level
\textit{rationale  density}  metric:\\ 

 \begin{math}
\textit{rationale density} = \frac{ \text{number of sentences labelled as \textit{Rationale}} }{\text{total number of sentences in a commit} }\\
\end{math}

\textit{Rationale density} is thus the percentage of  sentences that contain rationale in a commit. We compute this metric for all the commits. Then, we compute the  $average \; rationale \; density$ as follows:\\

\begin{math}
\textit{average rationale density} = \frac{ \sum_{\text{commits}} \text{rationale \; density} }{\text{number of commits that contain rationale} }\\
\end{math}

Our data set has an \textit{average rationale density}  of 61.43\%, 
in other words rationale information is present in about 60\% of a commit message. 
These results suggest that Linux developers support their decisions with a lot of rationale information, expressing it in a rather detailed way. 
This finding might serve as a rule of thumb guideline  for writing commit messages.

\vspace{0.1cm}
\setlength{\fboxsep}{10pt}%
\ovalbox{%
\begin{minipage}{7.4cm}
\textbf{Result 1 -- Presence of rationale:} 
Commit messages almost always contain rationale information. On average, around 60\% of the message contains rationale information. 
\end{minipage}}

\subsubsection{Factors impacting rationale.}
\label{sec:rationale_dependencies}

Here, we present analyses about the possible dependencies between the size of the commit (\textbf{RQ3}) and the  developers experience (\textbf{RQ4}), and  the \textit{rationale density}.

\textbf{RQ3. Does the quantity of rationale reported depend on the
commit message size?} The commit message size refers to the number of sentence entries obtained after preprocessing the commit message. To answer RQ3, first, we do a normality test~\cite{normal_tests} 
 on the distribution of the \textit{rationale density} of the commits of our dataset. Results indicate a non-normal distribution ($p\_value=0.02 < 0.05$). The normality test also indicates a non-normal distribution for the commit messages sizes ($p\_value=1.73e^{-55} < 0.05$). Since both distributions are not normal, we use Spearman's rank correlation coefficient~\cite{myers2004spearman} 
 to discover whether or not there is a correlation between the commit message size and its \textit{rationale density}. Results indicate that there is no correlation between them ($p\_value=2.45e^{-10}  < 0.05, R = -0.32$).

\begin{figure}
    \centering
    \includegraphics[width=0.33\textwidth]{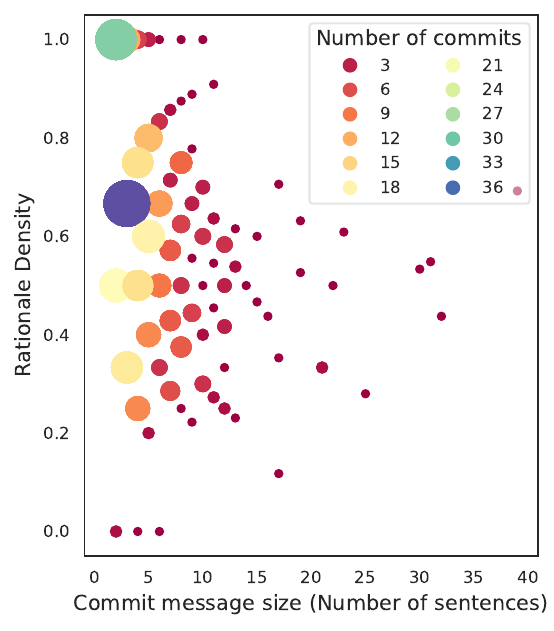}
    \caption{Commit message size versus rationale density}
    \label{fig:rationale_density_over_size}
    \vspace{-0.15in}
\end{figure}

In Figure~\ref{fig:rationale_density_over_size}, we show the \textit{rationale density} values versus the commit message size (i.e, number of sentences in a commit). The figure shows that the majority of the commits have fewer than 15 sentences, and that a lot of the short commits (fewer than 6 sentences) have a high \textit{rationale density} ( $>0.6$). The figure also shows that as a commit becomes longer, it tends to have between 40\% to 60\% of its sentences containing rationale information.

\textbf{RQ4. Does the quantity of rationale reported depend on the
developer experience?}
We consider the number of commits authored an indication of the developer's experience. We count the number of commits per author, as well as the average \textit{rationale density}  per author (i.e, we compute the mean of the \textit{rationale density} of the commits of each author). The normality test result indicates that the  number of commits per author is not a normal distribution ($p\_value=4.45e^{-33} < 0.05$). Spearman's rank correlation coefficient between  the number of commits per author and the average \textit{rationale density} per author indicates that the test is not significant ($ p\_value=0.635 > 0.05, R =-0.05$).

\begin{figure}
    \centering
    \includegraphics[width=0.33\textwidth]{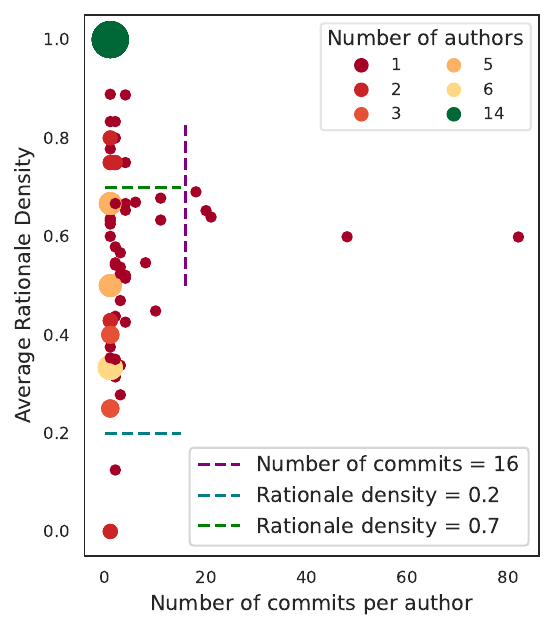}
    \caption{Commits per author vs average rationale density}
    \label{fig:rationale_density_per_author}
    \vspace{-0.15in}
\end{figure}

We visualize the average \textit{rationale density}   per author along with the number of commits per author in Figure~\ref{fig:rationale_density_per_author}. 
We identify four regions in the Figure: three for authors with few commits ($<16$) and one for authors with many commits ($>16$), separated by the vertical dashed line. In fact, only five developers have written more than 16 commits. All the other developers have written fewer than 16 commits, and most of them fewer than 10 commits. More experienced developers' commits have a consistent \textit{rationale density} near 60\%. This may indicate a guideline for the other developers to target. We further investigate these top contributors and their rationale densities in Section~\ref{sec:rationale_evolution}. 

\begin{figure*}
    \centering
    \includegraphics[width=0.7\textwidth]{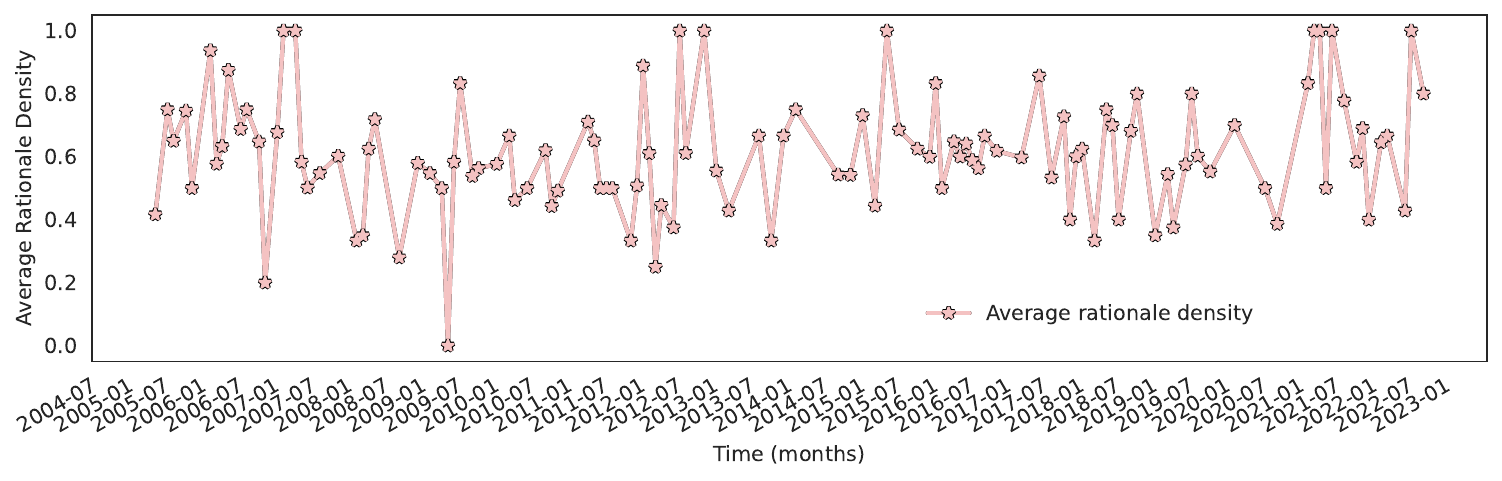}
    \caption{Monthly evolution of the average rationale density 
    }
\label{fig:rationale_evolution_monthly}
\end{figure*}
\begin{figure*}
    \centering
    \includegraphics[width=0.7\textwidth]{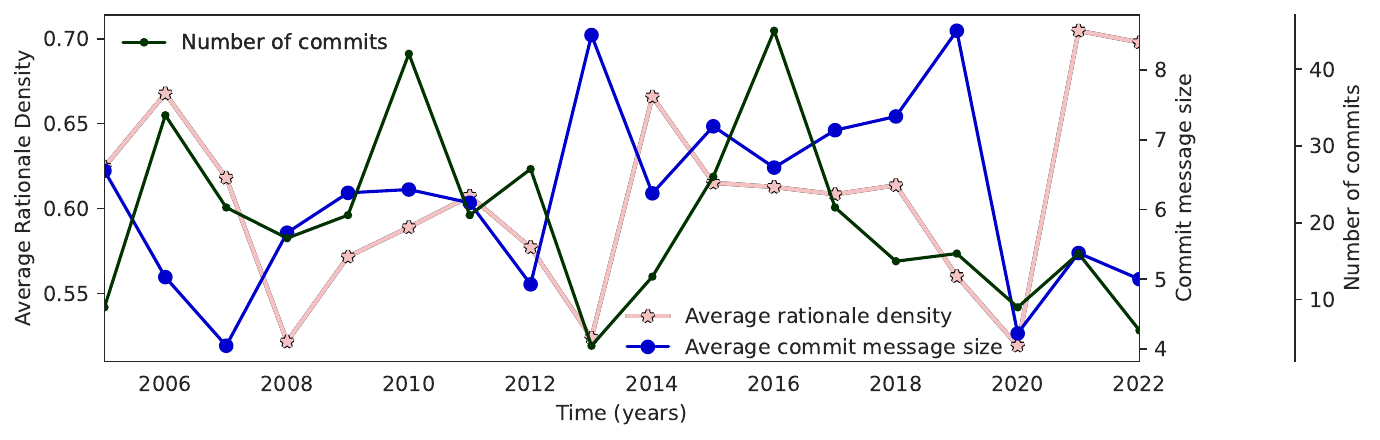}
    \caption{Yearly evolution of the average rationale density, the average commit message size and the number of commits }
    \label{fig:rationale_evolution}
\end{figure*}


In this analysis, we refer to the authors by their initials for ethical reasons~\cite{gold2020ethical}. The three regions for the  authors with few commits are separated by the two dashed horizontal lines pointing the densities of $0.2$ and $0.7$. Among these authors, three authors 
had a very low rationale density ($< 0.2$). Specifically, 
authors \textit{M.K.}  and \textit{P.E.}
had a rationale density of $0$ as each wrote only one commit that did not contain any rationale\footnote{\scriptsize \url{https://api.github.com/repos/torvalds/linux/git/commits/7b1915a989ea4d426d0fd98974ab80f30ef1d779}}\textsuperscript{,}\footnote{\scriptsize \url{https://api.github.com/repos/torvalds/linux/git/commits/c7ba5c9e8176704bfac0729875fa62798037584d}}. 
Author \textit{B.S.}
wrote two commits\footnote{\scriptsize \url{https://api.github.com/repos/torvalds/linux/git/commits/e222432bfa7dcf6ec008622a978c9f284ed5e3a9}}\textsuperscript{,}\footnote{\scriptsize \url{https://api.github.com/repos/torvalds/linux/git/commits/00f0b8259e48979c37212995d798f3fbd0374690}} but their rationale density was low ($0.125$) as one of them did not contain any rationale, and the other had only one sentence that contained rationale information. We note that all these commits are comparatively old (2009 or before) and that during the labelling process, we noticed that the quality of the messages improved over time. Finally, we note that these commits were approved for merging though they were lacking rationale, meaning that the message combined with the code change may have had  implicit rationale that was deemed sufficient. 

The top region shows that many authors with few commits have high \textit{rationale density}. Specifically, 14 authors had a density of $1$,  
e.g., 
author \textit{L.Z.}
wrote two commits\footnote{\scriptsize \url{https://api.github.com/repos/torvalds/linux/git/commits/97d87c9710bc6c5f2585fb9dc58f5bedbe996f10}}\textsuperscript{,}\footnote{\scriptsize \url{https://api.github.com/repos/torvalds/linux/git/commits/e115f2d89253490fb2dbf304b627f8d908df26f1}} with all  sentences containing rationale. 

\vspace{0.1cm}
\setlength{\fboxsep}{10pt}%
\ovalbox{%
\begin{minipage}{7.4cm}
\textbf{Result 2 -- Factors impacting rationale:} The quantity of rationale information  reported   depends neither on the commit message size nor the developers' experience. Experienced developers have a \textit{rationale density} around  60\%.
\end{minipage}}
\vspace{0.1cm}


\subsubsection{Evolution of rationale over time.}
\label{sec:rationale_evolution}

We first present the evolution of the rationale overall (\textbf{RQ5}) and then focus on the evolution of the rationale for the five main contributors (\textbf{RQ6}). We consider the authoring date of commits, not when it has been accepted.

\textbf{RQ5. How does rationale evolve over time? } To address this question, we visualize the evolution of the rationale density  over time.  First, we visualize the monthly evolution of the average \textit{rationale density} in Figure~\ref{fig:rationale_evolution_monthly}. The figure shows a consistency around $0.6$. To better contextualize this evolution, we show the yearly evolution of the average \textit{rationale density}, the average commit message size and the total number of commits in Figure~\ref{fig:rationale_evolution}. The figure shows great variations in the average commit message size and in the number of commits over the years. The figure does not suggest any relationship between these three variables. The correlation test between the average \textit{rationale density} and the number of commits confirmed this  ($ p\_value=0.63 > 0.05, R =-0.05$).

\begin{figure*}
    \centering
    \includegraphics[width=0.7\textwidth]{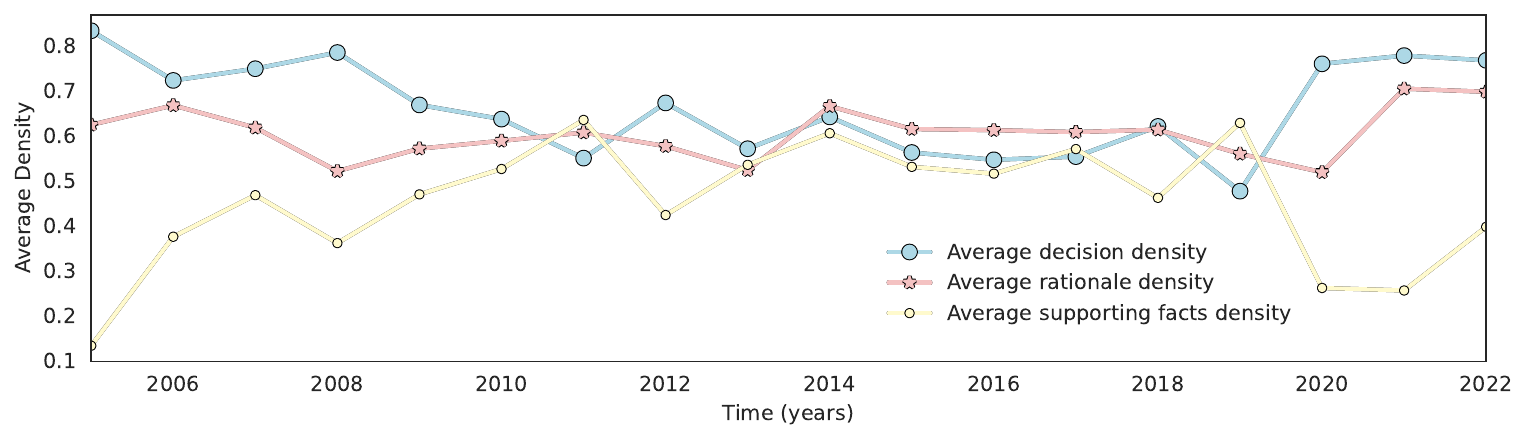}
    \caption{Yearly evolution of the average rationale density, the average decision density and the average supporting facts}
    \label{fig:evolution_decision_rationale_facts}
    \vspace{-0.1in}
\end{figure*}
\begin{figure*}
    \centering
    \includegraphics[width=0.7\textwidth]{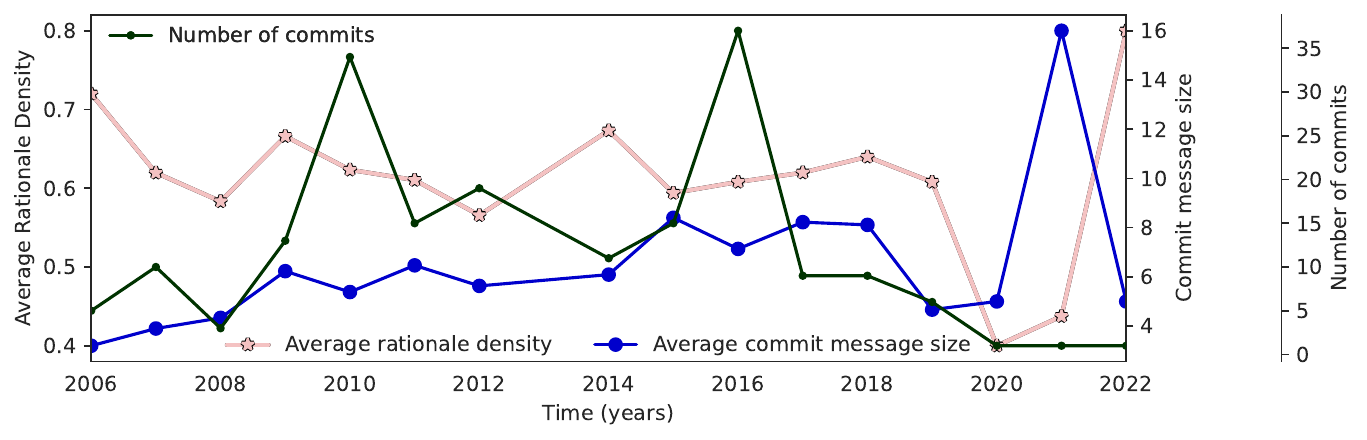}
    \caption{Evolution of average rationale density, average commit   message size, and number of commits for top 5 contributors}
    \label{fig:top_contributors}
    \vspace{-0.15in}
\end{figure*}

Figure~\ref{fig:evolution_decision_rationale_facts} shows the evolution of the average rationale density, the average \textit{decision density} and the average \textit{supporting fact density} per year. The \textit{decision density} and \textit{supporting facts density} are computed similarly to the \textit{rationale density}. 
We note that the \textit{decision density} is always high ($> 0.5$). However, the \textit{supporting facts density} is usually low ($<0.6$). We also note that in early and late years (2005-2010 and 2019-2022), the \textit{decision density} is slightly higher than the \textit{rationale density}, which is significantly higher than the \textit{supporting facts density}.  Between 2010 and 2019, the three densities seem to be close (around $0.55$), although the  \textit{decision density} and the \textit{rationale density} are almost always higher than the \textit{supporting facts density}.

\textbf{RQ6. How does rationale evolve over time for the five core contributors? }
The five main contributors in the OOM-Killer component are 
\textit{D.R.} (82 commits), \textit{M.H.} (48 commits), \textit{T.H.} (21 commits), \textit{K.M.} (20 commits), and \textit{O.N.} (18 commits). 
Together, these five contributors wrote 189 commits, i.e., around half of the 366 studied commits. This is consistent with past observations about open source projects, suggesting that a relatively small number of \textit{core} developers are responsible for most contributions~\cite{mannan2020relationship}. As discussed in Section~\ref{sec:rationale_dependencies}, these developers all had a consistent average overall \textit{rationale density} around 0.6. 
Figure~\ref{fig:top_contributors} shows the rationale density evolution, the average commit message size, and the number of commits per year for these contributors. Although \textit{rationale density} was consistently around 0.6 for all the years before 2020, it dropped to around 0.4 in 2020 and 2021 and went up to 0.8 in 2022. The figure also shows that the number of commits varies considerably each year, and that usually, the top contributors write short commits (fewer than 8 sentences), the year 2021 being the exception. The average commit message size (in terms of number of sentences) also seems to be trending up slightly over time.

\vspace{0.1cm}
\setlength{\fboxsep}{10pt}%
\ovalbox{%
\begin{minipage}{7.2cm}
\textbf{Result 3 -- Evolution of rationale over time:}  The level of \textit{Rationale density} remains consistent (around 0.6). \textit{Decision density} is always
high (> 0.5). \textit{Supporting facts density} is low(< 0.6). 
More experienced developers write short commit messages (fewer than eight sentences).
\end{minipage}}
\vspace{0.1cm}

Understanding the reasons behind this evolution would involve interviewing the developers and creating a deep understanding of the historical events of Linux Kernel development and culture as a whole (in case there are dependencies between the modules that impacted the OOM-Killer component). We consider performing this more in-depth  analysis as future work.

Investigating these impact factors is relevant for practitioners. For instance, if rationale trended down with experience or project age, a best practice might be to periodically train developers to provide more rationale. Instead, these results could suggest that the culture in the Linux kernel is sufficient to maintain the level of rationale, and should be encouraged in other software projects.

\subsubsection{Structure of commit messages.} 
Here, we discuss the average structure of a commit message. That is, what sentence category order developers prefer when elaborating their commit messages. 

\begin{figure}
    \centering  \includegraphics[width=0.40\textwidth]{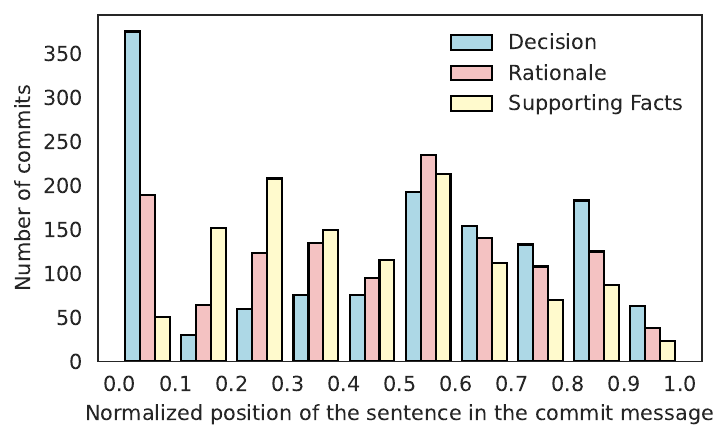}
    \caption{Category distribution over the normalized positions of commit message sentences  (considering overlap).}
    \label{fig:commit_strcuture}
\end{figure}

\textbf{RQ7. In what order do the categories mostly appear?}
We visualize the distribution of the identified categories over the normalized positions of the sentences of the commit messages in
Figure~\ref{fig:commit_strcuture}. The figure shows that, in the first 10\% of the commit message (the summary sentence), the \textit{Decision} category is present in over 350 commit messages.  
The \textit{Rationale} also appears frequently in this summary sentence. This could be explained by the overlap between the categories and especially by the value judgment words (e.g, ``fix'', ``simplify'', ``unused'', ...). e.g., the sentence  ``include cleanup: Update gfp.h and slab.h includes to prepare for breaking implicit slab.h inclusion from percpu.h''\footnote{\scriptsize\url{https://api.github.com/repos/torvalds/linux/git/commits/5a0e3ad6af8660be21ca98a971cd00f331318c05}} is labelled both \textit{Decision} and \textit{Rationale}.

In the first half of the commit (10\% to 50\%), the \textit{Supporting Facts} category is the most frequent, with the \textit{Rationale} category behind it. In the second half of the commit, the \textit{Rationale} category exceeds the \textit{Supporting Facts} category. That is, the \textit{Supporting Facts} are often found in the commit beginning,
and before the sentences containing rationale.  This can be explained by 
their reference to the past or current state of the system. The \textit{Supporting Facts} and the \textit{Rationale} categories seem to have a similar presence, especially in the middle of the commit (30\%-70\%). This can be explained by the substantial overlap between them (Figure~\ref{fig:dataset_categories}).   
By the end of the commit message (70\%-100\%), the \textit{Decision} category appears again, with the \textit{Rationale} category behind it.  Thus, the most common order seems to be : \textit{Decision} then \textit{Supporting Facts} then \textit{Rationale} then  \textit{Decision}. An example of this order is presented in Table~\ref{tab:coloured_commit_no_overlap}. 
The identified patterns may be used in the future as guidelines, or enforced via automated tools, to ensure well-structured commit messages, or suggest improvements before merging commits.

\vspace{0.1cm}
\setlength{\fboxsep}{10pt}%
\ovalbox{%
\begin{minipage}{7.2cm}
\textbf{Result 4  -- Structure of commit messages:} 
Developers tend to start and end their commit messages with \textit{Decisions}. \textit{Rationale} and \textit{Supporting Facts} appear in the middle of the commit, with \textit{Supporting Facts} usually preceding \textit{Rationale} sentences.
\end{minipage}}
\vspace{0.1cm}

We summarize the results of our analysis of the dataset in Table~\ref{tab:rq_summary}.

 \begin{table*}[bp]
    \footnotesize
      \caption{Research questions and their answers.}
    \begin{tabular}{p{2.1cm}|p{6cm}|p{8cm}}
        \textbf{Topic} &  \textbf{Research question}  &  \textbf{Answer} \\
         \hline
        \multirow{2}{*}{\shortstack{Presence of rationale}}  & RQ1.  How many commits contain rationale? & 98.9\% of the commits contain rationale. \\
        
    &   RQ2. How much of the commit contains rationale?  & Around 60\% of the commit message contains rationale information.\\
      \hline
\multirow{4}{*}{\shortstack{Factors impacting \\ rationale}} &RQ3. Does the quantity of rationale reported depend on
the commit message size?  & 
The quantity of rationale reported does not depend on
the commit message size (in terms of number of sentences). \\ 

&RQ4. Does the quantity of rationale reported depend on
the developer experience? & 
The quantity of rationale reported does not depend on
the developer experience. \\
\hline 

\multirow{5}{*}{\shortstack{Evolution of rationale \\ over time}} & RQ5. How does rationale evolve over time? &
The yearly evolution of rationale density is rather consistent around 0.5 and 0.7. The monthly evolution of rationale density however shows great fluctuation. \\

&RQ6. How does rationale evolve over time for the five core contributors?& 
The yearly evolution of rationale density is rather consistent around 0.6 for the five core contributors. This changes in the recent years 2020, 2021 and 2022 (with 1 commit per year).  \\ 

\hline 
\multirow{2}{*}{\shortstack{Structure of commit \\ messages}}   & RQ7. In what order do the categories mostly appear? & 
\textit{Decision} then \textit{Supporting Facts} then
\textit{Rationale} then \textit{Decision}. \\ 
&&

 \end{tabular} 
    \label{tab:rq_summary}

\end{table*}

\section{Threats to Validity}
\label{sec:threats}

\textbf{Construct validity.}
The measures we use to identify the presence of the different categories in the commits (i.e, the \textit{densities}), are quantitative measures. We do not measure this presence qualitatively. In fact,  we are not aware of any qualitative measures of rationale or decisions. Furthermore, we consider the number of commits written as an indicator of author experience. 
This is not necessarily true.
The number of commits is also computed as an overall classification for the author and not at the time of the commit. This introduces a threat as  authors might have gained experience during the studied period. However,
our findings mitigate this threat as we can see a clear difference between the authors with several commits and the authors with few commits (Figure~\ref{fig:rationale_density_per_author}). Thus, the data leads us to believe that the number of commits authored is a reasonable measure of experience. 

\textbf{Internal validity.}
It is possible that our manual labelling
process could have introduced unintentional bias, as only one annotator is a native English speaker. To mitigate
this, we had a total of three individuals involved in labelling the commits, developing a shared codebook to use as guide. We also had several piloting rounds and discussions throughout the labelling process. An average 
Fleiss kappa of ~0.65 indicates a good reliability for our labelling. To mitigate cases of potential bias due to annotator background, we took an inclusive labelling approach during consolidation meetings. This approach assigned a final set of labels to a sentence based on the union of labels from all annotators. 

Our decision to filter the 99 \textit{Inapplicable} sentences might  introduce bias, as the removed sentences might contain relevant information. To mitigate this threat, we only removed the sentences where two or more annotators agreed that it was \textit{Inapplicable}. 

We only studied commits that were approved by the Linux maintainers, which could introduce survivor bias. 
This is acceptable as we are interested in the rationale of the software as it is, not as it could have been. Comparing our findings with the characteristics of rationale in rejected commits is however an interesting research question for the future. 
Additionally, some metrics are calculated based on a low number of commits (e.g. some monthly and yearly averages when studying rationale evolution). This represents a threat as it may mean that the results are more metrics of a few particular commits rather than true evolutionary trends.

\textbf{External validity.} 
We only studied the commit history that is available on Git, from 2005 onwards. Our observations might not be applicable to contributions in the earlier years of Linux.

Additionally, 
there is no reason to assume that our findings can be extrapolated to a) other Linux components, or b) other OSS  projects.
Regarding a), we note that, to the best of our understanding, there is nothing radically different about the development of the OOM Killer module compared to the rest of the kernel. Regarding b), we note that while there are real differences between the way contributions are managed in Linux compared to other large OSS projects, there are also lessons to be learned across projects~\cite{bettenburg2015management}. We view this paper as a step towards creating a better understanding of rationale in OSS generally, which can lead to improvements in automatic rationale extraction across projects.

\section{Related Work}
\label{sec:related}

\noindent\textit{About the Linux Kernel.}
Spinellis  \textit{et al.} studied the evolution of Unix from an architectural perspective~\cite{spinellisEvolutionUnixSystem2021}. Patel \textit{et al.}  studied the logging practices in the Linux Kernel~\cite{patelSenseLoggingLinux2022}.
Trinkenreich \textit{et al.}~\cite{trinkenreich2023belong} surveyed Linux Kernel contributors to develop a theoretical model for the sense of virtual community in open source software. 
However, none of this prior research focused on the rationale in the kernel commit messages.

\noindent\textit{About Commit Messages.} Al Omar \textit{et al.}~\cite{alomarCanRefactoringBe2019} explored how developers document their activities in refactoring commits. They manually extracted the patterns used when refactoring (i.e, the keywords or phrases that frequently occur in refactoring-related commits). 
Other researchers tried to detect developer rationale  from the refactoring commit message to recommend the refactoring that would meet the
developer’s intentions~\cite{rebai2020recommending}. Our work is different as we do not focus on one specific category of commits (i.e, refactoring commits). They also consider the rationale as the difference of the  quality attribute values before and after the commit, while we study the rationale information present in free text.

Approaches  have been proposed to automatically generate commit messages~\cite{tao2022large}.
However, these are based on datasets that include poorly phrased commit messages. In fact, it is  only recently that researchers have studied commit message quality~\cite{tian2022makes,li2023commit}. 
Tian et al. 
has tried to define what constitutes a “good”
commit message~\cite{tian2022makes}, and has found that it should summarize \textit{what}
was changed, and describe \textit{why} those changes are needed. 
They also developed a taxonomy based on recurring
patterns in commit message expressions and   proposed a good-message
identification tool.  
Li \textit{et al.}~\cite{li2023commit} consider link contents in addition to the commit message to train classifiers for the automatic identification of good commit messages. They also studied the commit quality evolution (i.e, whether a commit contains \textit{why} and \textit{what} information).  
Similar to our study, these works (i.e, \cite{li2023commit,tian2022makes}) involved  manually creating  a commit dataset from open source projects and investigating the temporal evolution aspect. They also distinguished the evolution for the core and non-core developers. However, they only considered the evolution of the existence of \textit{what} (decision) and \textit{why} (rationale) information  over time, while we study the evolution of their quantities. Interestingly, our results indicate stable 
amounts, while they found that the overall quality degrades over time. In addition, they study the correlation between defect proneness and the quality of the  commit message as well as  the quality of prior commits, while we focus on different correlations.  They have performed their studies at the commit-level, while we label and analyze at the sentence-level. Finally, our study focuses on a component from the Linux kernel, while they report results across 32 Apache projects. 

\noindent\textit{About Design Decisions and Rationale.}
Li \textit{et al.} created a ground
truth dataset from the Hibernate developer mailing list by labelling sentences as \textit{decision} or \textit{non-decision}~\cite{li2020automatic}. 
Bhat \textit{et al.} manually analyzed and labelled
more than 1,500 issues from two large open source repositories~\cite{bhatAutomaticExtractionDesign2017}. They classified the issues into different decision categories. Unlike our work, they only considered decisions and discarded rationale expressions.

%
Sharma \textit{et al.} studied Python email  archives and produced a labelled dataset of rationale sentences  behind acceptance of Python Enhancement Proposals (PEP)s~\cite{sharmaExtractingRationaleOpen2021}. They also  defined 11 categories of rationale and used them to classify the rationale sentences.
This work differs because a) they consider emails while we consider code commits and b) their work is specific to the PEPs, as the decisions they consider refer to the possible transitions between the PEP states (e.g, draft, accepted, refused).
Kleebaum  \textit{et al.} propose Condec tools to support documenting and managing decision knowledge for change impact analysis~\cite{kleebaum2021continuous}.  These tools include an automatic text classification feature for rationale. To provide ground truth for the classifier, the authors  manually labelled  the textual artifacts they produced when developing the Condec tools (Jira issue text, commits and code comments) to whether or not it contains rationale and they identified rationale categories (decision, issue, alternative, pro-argument, con-argument). 
Different from our work, they consider all textual artifacts and not only code commits and do not analyze the resulting dataset. 

Vanderven  \textit{et al.} created a commit dataset from 710 Git projects, by asking six experts to label 100 commits if they involved a decision, rationale for a decision, and decision alternative information~\cite{vandervenMakingRightDecision2013}. However, their dataset is not publicly available, and they work at the commit-level. 
Furthermore, their analysis only investigated the presence of rationale, decisions and alternatives in the commit messages. They neither studied the evolution of their quantities over time, nor the order in which they appeared.
%
Alkadhi  \textit{et al.} investigated the presence of rationale in the Internet Relay Chat (IRC) chat messages of three OSS projects: Apache Lucene, Mozilla Thunderbird and Ubuntu~\cite{alkadhiHowDevelopersDiscuss2018}.
They manually indicated if each message contains rationale, and  identified the type of rationale elements (decision, issue, alternative, pro-argument, con-argument) included in the message.   They also provided evidence of rationale existence in the chat messages, and the frequency of different rationale elements. Furthermore, they
explored which developers contribute to rationale in the messages. This work differs from ours because 1) they consider chat messages instead of code commits, 2) they work at the message-level while we work at the sentence level, and 3) they did not consider the Linux kernel project.

Recently, several researchers proposed rationale representations.
  Hesse  \textit{et al.} proposed
a documentation model for decision knowledge built upon
the results investigating the comments to 260 issue reports
from the Firefox project~\cite{hesse2020supporting}. Soliman  \textit{et al.} worked on
an empirically-grounded ontology for architecture knowledge
from StackOverflow posts~\cite{soliman2017developing}. 
Neither one of these works considered rationale representation for code commits. 
Alsafwan  \textit{et al.}
 performed interviews and surveys with 
developers to study their perspective of rationale for code
commits, and found that they decompose the rationale of code commits into 15 separate components~\cite{alsafwanDevelopersNeedRationale2022}. 

\section{Conclusion and Future Work}
\label{sec:conclusion}

We explored the presence of rationale in the commit message history of the Linux OMM-Killer. We created a dataset by extracting and manually labelling the commits. We analyzed it over seven research questions about the presence of rationale, its evolution, the factors impacting it and the structure of rationale information. The results are summarized in Table~\ref{tab:rq_summary}.

This lead us to insights about the nature of rationale. First,
it is prevalent in the commit messages with a stable amount  over the studied period. Second, there is a substantial overlap between the three defined categories, as expressions of rationale are very often interwoven with decisions and supporting facts. This may suggest the need to consider a finer-grained analysis in the future (e.g, clause-level analysis). Third, 
experienced developers are responsible for writing almost half of the commits, have a consistent rationale density, and usually write short messages. Finally, a common commit structure seems to emerge: \textit{decision} then \textit{supporting facts} then \textit{rationale} then \textit{decision}.






In the future, first we aim to increase the dataset quality and richness. We plan to classify the commits (e.g, trivial fix, refactoring, new feature) to investigate how rationale varies according to context. 
We are also investigating adding a \textit{large language model} as another annotator. Syriani \textit{et al.} show that ChatGPT can offer human-level performance for a similar labelling task, but this requires systematic prompting and performance assessment~\cite{syriani2023assessing}.
Second, we will investigate \textit{rationale management}, e.g., training an automatic  rationale extraction model, using our dataset as the ground truth. Along with our rationale structure insights, we could automatically detect commit messages that lack rationale or lack the correct structure when commit messages are written or requested to be merged.






\begin{acks}
This work is partially funded by a Fonds de Recherche du Québec – Nature et Technologies (FRQNT) 
Doctoral Research Scholarship (B2X).
\end{acks}

\bibliographystyle{ACM-Reference-Format}
\bibliography{rationale, linux}

\end{document}